\documentclass[trackchanges,twocolumn,tighten]{aastex7} % Concelled Linenumber
\usepackage{amsmath}
\usepackage{newtxtext,newtxmath}
\usepackage{makecell}

\begin{document}

\title{Plasma lensing modeling of substructures on pulsar scintillation screens}%Template \aastex v7 Article with Examples\footnote{Footnotes can be added to titles}}

\author[orcid=0000-0002-2575-0241,sname='Xu']{Zhu Xu}
%\altaffiliation{Kitt Peak National Observatory}
\affiliation{South-Western Institute for Astronomy Research (SWIFAR), Yunnan University, 650500 Kunming, P. R. China}
\email[show]{zhu@mail.ynu.edu.cn} 

\author[orcid=0000-0003-2076-4510,sname='Shi']{Xun Shi}
%\altaffiliation{Kitt Peak National Observatory}
\affiliation{South-Western Institute for Astronomy Research (SWIFAR), Yunnan University, 650500 Kunming, P. R. China}
\email[show]{xun@ynu.edu.cn} 
\begin{abstract}

Radio pulsars, as highly coherent point sources, serve as powerful probes of the ionized interstellar medium (IISM). Pulsar scintillation observations have revealed inverted arclets on the secondary spectrum, indicating quasilinearly aligned images created by substructures on a scintillation screen. The density profiles of these substructures remain unconstrained but are crucial to identifying their physical nature. This work employs a plasma lensing framework to study observable features of substructure phase screens. Using three lens models, we identify the substructure properties that can be constrained by observables. The outer caustic is the most prominent feature of a lensing substructure, measurable via multiepoch or ultrawideband observations. Its location constrains the maximum column density gradient of the substructure. The inner caustic, though difficult to observe except for substructures capable of producing extreme-scattering events, directly indicates the substructure size. Even when caustic locations are not observed, the minimum span where substructure images exist can be measured and used to place a lower limit on the column density amplitude. The logarithmic brightness of individual arclets forms a concave function of the pulsar-lens angular separation, contrasting with the convex brightness distribution of all substructure images—highlighting the complementarity of individual arclets to statistical studies. These findings reveal the potential of pulsar scintillation to uncover IISM substructure and underscore the need for multiepoch and/or ultrawideband measurements to constrain discrete lensing morphologies and help reveal the nature of interstellar plasma structures.
\end{abstract}



\section{Introduction} 

Coherent density structures within the ionized interstellar medium (IISM) act as plasma lenses, deflecting radio waves that pass through them and causing interference among coherent wavefronts. This effect leads to the scintillation of signals from compact radio sources, such as pulsars and fast radio bursts (FRBs), %quasars and pulsars, 
in a manner analogous to the optical twinkling of stars. The dynamic spectrum %(DS) 
serves as the predominant method for recording pulsar scintillation data, capturing variations in brightness across both time and frequency \citep{rickett1969frequency,gupta1994refractive,cordes1986multiple,johnston1998scintillation}. The dynamic spectrum offers critical insights into the propagation characteristics of pulsar radio signals through the interstellar medium.

The two-dimensional power spectrum of the dynamic spectrum%(DS)
, known as the 
``secondary spectrum''
%"secondary spectrum"%(SS)
, is employed to analyze periodic fringes occasionally observed in the dynamic spectrum \citep{ewing1970observations,roberts1982dynamic,hewish1985quasiperiodic,cordes1986multiple,wolszczan1987interstellar}. These fringes manifest as discrete power concentrations in the secondary spectrum, typically attributed to interference effects arising from the coherent nature of signals emitted by compact sources such as pulsars.

Since the early 2000s, distinct parabolic arcs have been observed in the secondary spectra of pulsars \citep{stinebring2001faint}, revealing a quadratic relationship between the differential Doppler shift ${\rm f_D}$ and the delay $\tau$ \citep{walker2004interpretation,cordes2006theory}. In some cases, multiple parabolic arcs are observed, suggesting the presence of multiple scattering screens at different distances along the line of sight \citep{putney2006multiple,stinebring2022scintillation,main2023thousand,ocker2024pulsar,reardon2025bow}.

Inverted arclets have been observed in the secondary spectra of certain pulsars, with their apices lying on the main parabola and sharing the same curvature \citep{hill2003pulsar,hill2005deflection,brisken2010100,sprenger2022doublelens}. These features result from the interference between different scattered rays of the pulsar signal, which are deflected by discrete structures within the %ionized interstellar medium (IISM) 
IISM that act as the plasma lenses. Due to the relative motion between the pulsar, the plasma lensing structure, %lens,
and the observer, these arclets shift along the main parabolic arc over time. Observations by \citet{hill2003pulsar} and \citet{hill2005deflection} indicate that these arclets can persist for over a month, suggesting the stable existence of the plasma lenses.

Using a VLBI observation of the scintillation of PSR B0834+06, \citet{brisken2010100} reconstructed the scattered images and found them lining up along a thin line that crosses the line of sight. The linear alignment of the scattered images has been confirmed by later reconstructions of the scintillation of PSR B0834+06 \citep{simard2019disentangling,zhu2023pulsar} and has also been suggested by the observations of PSR B1508+55 \citep{sprenger2022doublelens}. This implies that the density structures of the underlying scintillation screens are in the form of either parallel filaments or threaded beads, but more likely parallel filaments, given that large apex offsets have not been observed \citep{shi2021morphology}. The currently most plausible physical explanation for such a geometry is corrugated reconnection sheets \citep{pen2012refractive,pen2014pulsar,liu2016pulsar,simard2018predicting}.

In addition to inverted arclets in the secondary spectrum% (SS)
, extreme scattering events (ESEs)---characterized by significant flux density variations over days to weeks and primarily attributed to the IISM %ionized interstellar medium (IISM) 
\citep{fiedler1987extreme,fiedler1994summary,lazio2001dualfrequency,bannister2016realtime,clegg1998gaussian}---also indicate the presence of small-scale structures within the IISM acting as plasma lenses. %Furthermore, 
\citet{jow2024cusp} proposed the cusp lens model, an extension of the fold lens framework \citep{pen2014pulsar,simard2018predicting}, within the ``doubly catastrophic framework''. 
%"doubly catastrophic framework." 
%This model offers a unified description of scintillation and ESEs as manifestations of the same phenomenon,
This model extends the foundational framework of \citet{romani1987radio} by offering a unified description of scintillation and extreme scattering events (ESEs) as manifestations of the same phenomenon, employing catastrophic lenses defined by a simple potential and a minimal set of parameters, observable along lines of sight intersecting corrugated reconnection sheets.

%\zhu{The astrophysical origin of these plasma lenses remains a great mystery.}
While plasma lenses have been observed in specific environments such as ionized filaments in the Crab Nebula, the nature and properties of the plasma structures responsible for lensing events in the general ISM remain poorly constrained.
Given their tiny sizes on AU scales, other ISM observations can hardly offer a clue. Currently, the most straightforward way to make progress is to measure the detailed properties of inverted arclets in pulsar scintillation secondary power spectra %SS 
and to link these measurements to the physical properties of the underlying plasma lenses. This requires modeling of the lensing properties of various plasma lens shapes, which is what we aim to study in this paper.

\section{Lens modeling of pulsar scintillation secondary spectrum}

%\subsection{Justification for geometric lensing}
\subsection{Eikonal approximation of pulsar scintillation secondary spectrum}
Astrophysical lensing refers to the phenomenon in which signals from a source are deflected by intervening 
lenses, e.g., gravitational lenses or plasma lenses
%potentials—such as those responsible for plasma lensing or gravitational lensing—
between the source and the observer %within our universe
. 
%Traditionally, studies of astrophysical lensing have relied on the ray optics framework. However, as the lensing of coherent astrophysical sources, including pulsars, fast radio bursts, and gravitational waves, becomes observationally significant, wave effects can no longer be overlooked. To accurately describe the lensing of such coherent sources, it is necessary to employ the Kirchhoff-Fresnel integral 
%\zhu{A general framework to describe astrophysical lensing by a single lens is the Kirchhoff-Fresnel integral, which gives the wave amplitude $E$ received by an observer for a point source with a unit magnitude:}
Astrophysical lensing by a single lens can be described using the thin-screen approximation, in which the lens is modeled as a phase-changing screen located at a single plane. The wave amplitude $E$ received by an observer is then given by the Kirchhoff-Fresnel integral \citep[e.g.,][]{born1999principles} over this plane for a point source with a unit magnitude:
\citep{coles1987refractive,cordes1986refractive,shi2024acquiring,shi2024stokes,shi2024lensing,jow2022regimes}, %potentially supplemented by higher-order asymptotic approximations.
\begin{equation}
 	%\boldsymbol{E}
      E(\boldsymbol{\beta},\nu)=\frac{1}{2\uppi \rm{i} \theta_{F}^{2}}\int{{\rm d}^{2}\boldsymbol{\theta}{\rm exp}[\rm{i}\Delta \Phi(\boldsymbol{\beta,\theta},\nu)]} \,.
 	\label{eq:Kirchhoff-Fresnel integral}
\end{equation}
%In this formulation, the lens is assumed to be at a fixed position, with $\boldsymbol{\theta}$ representing the angular coordinate on the lens plane. The relative motion among the source, lens, and observer is fully encapsulated by the angular coordinate of the source, $\boldsymbol{\beta}$, which incorporates time dependence. 
The total phase change along the propagation path, $\Delta\Phi$
, is explicitly given by:
\begin{equation}
    \Delta \Phi(\boldsymbol{b}) = \frac{2\pi D_{\rm eff}}{\lambda} \phi(\boldsymbol{b}) + \pi \frac{|\boldsymbol{b}|^2}{\lambda D_{\rm eff}},
\end{equation}
where $\phi(\boldsymbol{b})$
is the phase screen that describes the phase change induced by the plasma at impact parameter $\boldsymbol{b}$. For a ionized plasma screen, $\phi$ is related to the electron column density $N_e$ by:
\begin{equation}
    \phi(\boldsymbol{b}) = \frac{r_e \lambda}{2\pi} N_e(\boldsymbol{b}),
\end{equation}
where $r_e$ is the classical electron radius. The second term in Equation (X) represents the geometric phase delay due to Fresnel diffraction.

Here, $\nu$ is the observing frequency, $\theta_{\rm F}$ is the angular Fresnel scale, and $\boldsymbol{\theta}$ and $\boldsymbol{\beta}$ are the angular coordinates of the lens and the source, respectively. % The Fresnel scale $\theta_{\rm F} = $...}
%\zhu{The total phase delay caused by lensing}
The total phase delay, $\Delta\Phi$, %comprises two primary components for high precision. 
%The phase delay 
is explicitly given by:
\begin{equation}
 \Delta\Phi = \frac{2\uppi \nu}{c}D_{\rm eff}\frac{|\boldsymbol{\theta}-\boldsymbol{\beta}|^{2}}{2} - \nu\tau_{0}\psi= \nu\tau_{0}\left(\frac{|\boldsymbol{\theta}-\boldsymbol{\beta}|^{2}}{2} - \psi\right)
 \label{eq: phase delay}
\end{equation}
where $\tau_0 = 2\uppi D_{\rm eff} / c$, $D_{\rm eff}$ denotes the effective distance, determined by the relative distances among the pulsar, the scintillation screen, and the observer, and $\psi$ is its lensing potential that is proportional to the column density difference $\Delta n_{\rm e}$ of the substructure lens,
$\psi \propto \Delta N_{\rm e} \nu^{-2}$ \citep[e.g.,][]{shi2021plasma}.
Of the two components of the total phase delay, the first 
is the geometric phase delay, resulting from wavefront distortion caused by the plasma lens. The second
arises from the lens potential; in the case of plasma lensing, this manifests as a phase shift of the wavefront induced by fluctuations in electron number density. %The second is the geometric phase delay, resulting from wavefront distortion caused by the plasma lens.

\begin{figure}[ht!]
\centering
    \includegraphics[width=0.9\columnwidth]{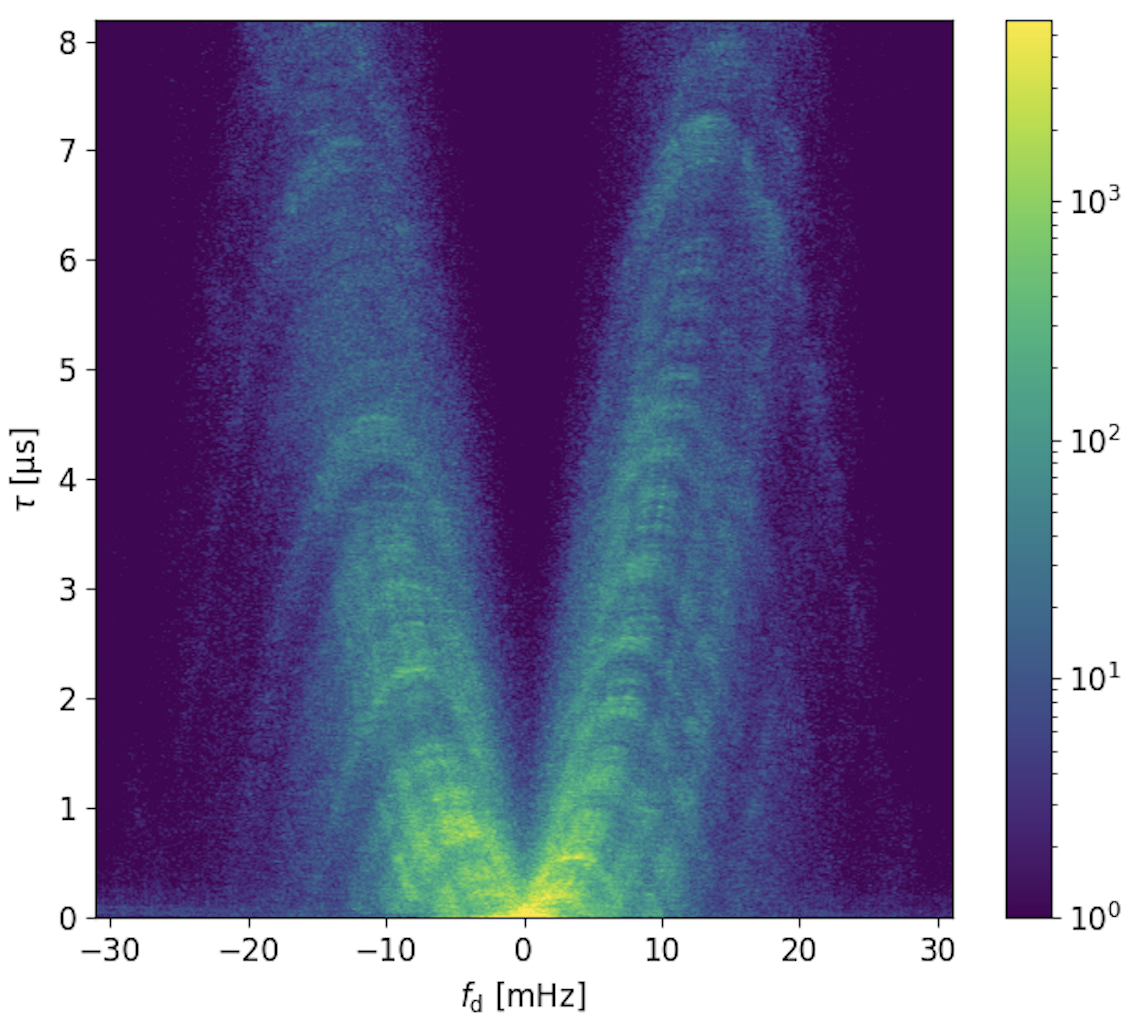}
    \caption{Secondary spectrum of pulsar B1508+55 observed by the FAST telescope. Along the main parabolic arc are numerous inverted arclets, each corresponding to an image created by a lens on the scintillation screen.} 
    \label{fig:sec}
\end{figure}

%\subsection{Pulsar scintillation secondary spectrum}
For a plasma lens to create a clear inverted arclet in the secondary spectrum of a pulsar, or to create an ESE, its lensing potential must be large, i.e., corresponding to a strong lens. According to \citet{shi2024lensing}, the lensing effects by such a lens can be well-captured by the eikonal approximation, that is, the interference between a discrete set of lensing images \citep[e.g.,][]{walker2005electric},
%Under the electric field representation introduced by %\citet{walker2005electric} for pulsar intensity spectra, the received electric field $E$ of a pulsar is expressed as the sum of contributions from all images produced by the scintillation screen. Up to an arbitrary normalization, this is given by:
\begin{equation}
 	E(\nu,t) = \sum_{j}\sqrt{\mu_{j}}{\rm exp}\left[2\uppi {\rm i}\left(f_{\rm{d},j}(t-t_{0})-\tau_{j}(\nu-\nu_{0})\right)\right]
\end{equation}
where %$\phi_{j} = \tau_{j} (\nu - \nu_{0})$ represents the phase of image $j$
$\mu_j$, $f_{\rm{d},j}$, and $\tau_j$ are the magnification, Doppler shift, and delay of image $j$
, and $t_0$ and $\nu_{0}$ denote the central time and frequency of the observation, respectively. %To characterize the electric field of an image $j$, 
The Doppler shift is defined as:
\begin{equation}
 	f_{dj} = \frac{\partial\Delta\Phi_{j}(\nu,t)}{2\uppi\partial t} = \frac{\nu}{c}V_{\rm eff}|\boldsymbol{\theta}_{j}-\boldsymbol{\beta}| 
\end{equation}
and the delay as:
\begin{equation}
 	\tau_{j} = \frac{\partial\Delta\Phi_{j}(\nu,t)}{2\uppi\partial \nu} = \frac{D_{\rm eff}|\boldsymbol{\theta}_{j}-\boldsymbol{\beta}|^{2}}{2c} \,.
\end{equation}
%with the amplitude $\sqrt{\mu_{j}}$ evaluated at $t = t_{0}$ and $\nu = \nu_{0}$, where $\mu_{j}$ is the magnification of image $j$.
%For an intervening plasma lens located at a fractional distance $s$ from the pulsar to the observer, the effective distance is defined as $D_{\rm eff} \equiv D (1 - s) / s$, where $D$ is the pulsar distance, and the effective velocity as $\boldsymbol{V}{\rm eff} \equiv \boldsymbol{V}{\rm sr} / s$, where $\boldsymbol{V}_{\rm sr}$ is the relative velocity of the plasma lens along the pulsar-observer line of sight (see, e.g., \citet{brisken2010100}).

The pulsar intensity, %defined as 
$I = EE^*$, %varies with time and frequency and is recorded as a dynamic spectrum. This 
is expressed as:
\begin{equation}
 	I(\nu, t) = \sum_{j,k} \sqrt{\mu_{j}\mu_{k}} {\exp}\left[ 2\uppi {\rm i}\left((f_{\rm{d},j} - f_{\rm{d},k})(t-t_{0}) - (\tau_{j} - \tau_k)(\nu-\nu_{0})\right) \right] \,.
    \label{eq: dynamic spectrum}
\end{equation}
%where $\xi_{jk} := \xi_{j} - \xi_{k}$, with $\xi$ representing either ${\rm f_{D}}$ (Doppler shift) or $\tau$ (delay). The pattern in the dynamic spectrum arises from the interference of image pairs.

The corresponding secondary spectrum, defined as $S({\rm f_{\rm{d}}}, \tau) = \tilde{I} \tilde{I}^{*}$, is given by:
\begin{equation}
 	S( f_{\rm{d}},\tau) = \sum_{j,k}\mu_{j}\mu_{k}\delta_{D}( f_{d} - f_{d,j} + f_{\rm{d},k})\delta_{D}(\tau + \tau_{j} - \tau_{k})
    \label{eq: secondary spectrum}
\end{equation}
where $\delta_{D}$ denotes the Dirac delta function. According to Equation~\ref{eq: secondary spectrum}, the secondary spectrum is computed by summing contributions from all image pairs. Each pair $(j, k)$ contributes power proportional to $%A_{j}^{2} A_{k}^{2} \propto 
\mu_{j} \mu_{k}$ at the position $(f_{\rm{d}, j} - f_{\rm{d}, k}, -\tau_{j} + \tau_k)$ in the $({\rm f_{\rm{d}}}, \tau)$ plane.

\begin{figure}[ht!]
    \includegraphics[width=\columnwidth]{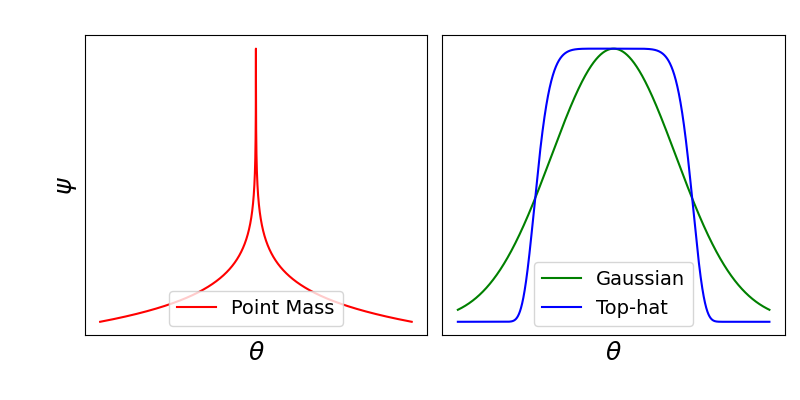}
    \includegraphics[width=\columnwidth]{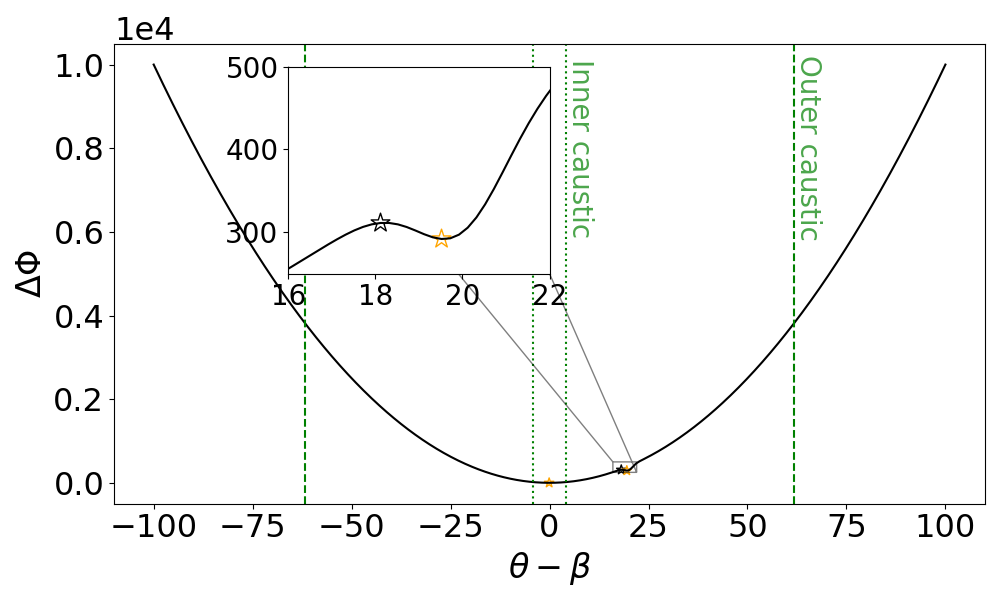}
    \caption{\textbf{Illustration of modeled lens shapes and the phase delay $\Delta\Phi$}.
    Very different shapes of the lens models (upper panel) are chosen to represent distinct astrophysical origins. 
    The total phase delay of the Gaussian lens (lower panel) includes a parabolic-shaped geometrical delay (grey line) apart from the dispersive delay caused by the lens. The images (stars) were generated on the phase delay; the image at the zero point of the x-axis is the source itself, and the extra images are created by the lens. The dashed green lines are caustics for the Gaussian lens (green curve). Only when the lens is between inner and outer caustics, two new extremes will be created on the phase delay, which means two new images will be created. 
    }
    \label{fig:model_shapes}
\end{figure}

Figure~\ref{fig:sec} shows a typical secondary spectrum of pulsar ${\rm B1508+55}$ observed by the Five-hundred-meter Aperture Spherical radio Telescope (FAST). For this strongly scattered pulsar, numerous inverted arclets are distributed %regularly arranged 
along the main parabolic arc. % of pulsar ${\rm B1508+55}$ observed by the FAST telescope. 
Each arclet corresponds to an image formed at a stationary-phase point (a lensing point) on the scintillation screen. %, the extra images created by the plasma lens being located a few magnitudes around the lens, thus 
In most cases, the brightest, direct image of the pulsar is located approximately at an angular location of the source, i.e., $\boldsymbol{\theta} \approx \boldsymbol{\beta}$. Thus, it has the minimum delay among all images, as well as a vanishing Doppler value. For each of the extra images created by the plasma lenses, its angular location with respect to the source location, $\boldsymbol{\theta}-\boldsymbol{\beta}$, is proportional to the Doppler value of the corresponding arclet apex.
%the arclet position on doppler axis is related to the relative lens location to the pulsar-observer line-of-sight, %so in the lens-observer coordinate, the location of pulsar 
%$|\boldsymbol{\theta}-\boldsymbol{\beta}| \propto {f_{\rm{d}}}$.

\subsection{Lens equation}
Theoretically, the angular location of the lensed images can be computed using
%Under 
the stationary phase condition $\nabla_{\boldsymbol{\theta}}\Delta\Phi = 0$. %for the total phase delay Eq.~\ref{eq: phase delay}, we have
%\begin{equation}
% 	\nu\tau_{0}(\boldsymbol{\theta} - \boldsymbol{\beta} - %\boldsymbol{\alpha})=0
% 	\label{eq: stationary phase condition}
%\end{equation}
%then the lens equation maps the source plane coordinate $\beta$ to the image/lens plane coordinate $\theta$ is
Inserting the expression for the total phase delay (Eq.~\ref{eq: phase delay}), the stationary phase condition leads to the lens equation \citep[e.g.,][]{schneider1992gravitational}
 \begin{equation}
 	\boldsymbol{\beta} = \boldsymbol{\theta} - \boldsymbol{\alpha}
 	\label{eq:lensEq}
 \end{equation}
 with the deflection angle 
 \begin{equation}
 	\boldsymbol{\alpha}(\boldsymbol{\theta}) \equiv \nabla \psi(\theta) \,.
 	\label{eq:Deflection Angle}
 \end{equation}
 %$\psi(\theta)$ is the potential of the lens, we can build a lens model by constructing their lens potential like what we do in section~\ref{sec:Lens Models}.

As lensing preserves surface brightness, 
the magnification $\mu$ of an image due to lensing %, which preserves surface brightness, 
is determined by the ratio of the surface areas post-lensing to pre-lensing, or equivalently, the Jacobian determinant of the lens mapping, 
 
\begin{equation}
    \mathcal A = \frac{\partial \boldsymbol{\beta}}{\partial \boldsymbol{\theta}}  = 1 -\partial_{\theta}\boldsymbol{\alpha}
    =1 - \partial_{\theta}^{2}\psi
    \label{eq:Jacobian}
\end{equation}
as
%The brightness of an image at position $\theta$ is thus modified by the magnitude of  $|\mu(\theta)|$ where  
\begin{equation}
	\mu=\frac{1}{|{\rm det}\mathcal A|} \,.
	\label{eq:Magnification}
\end{equation}

The regions on the source plane where $\mu$ becomes infinite are termed caustics. %These caustics demarcate the boundaries where the number of lensed images shifts by two. 
As a source transits a caustic, two new images appear or two existing ones are extinguished. The theoretical infinite magnification at caustics reflects an imperfection in the eikonal approximation and does not truly occur in practical scenarios, even for a point source.

\subsection{Lens Models}
\label{sec:Lens Models}

Currently, we know little about the shape of the plasma lenses in the IISM that create the inverted arclets, apart from their often highly anisotropic %i.e. nearly one-dimensional 
geometry, so we use 1D lens models for simplicity and practicality.
Here we adopt three lens models with distinctly different lens shapes to mimic different %cover the typical 
astrophysical origins: Gaussian, %step-like,
top-hat and spiky (point mass). %Exploring their differences and similarities under the lensing framework.
By exploring the difference of their lensing properties, we expect to get a clue of how well pulsar scintillation observations can constrain the geometry of underlying plasma lenses.

% Caustic vs Amplitude
\begin{figure}[ht!]
	\includegraphics[width=\columnwidth]{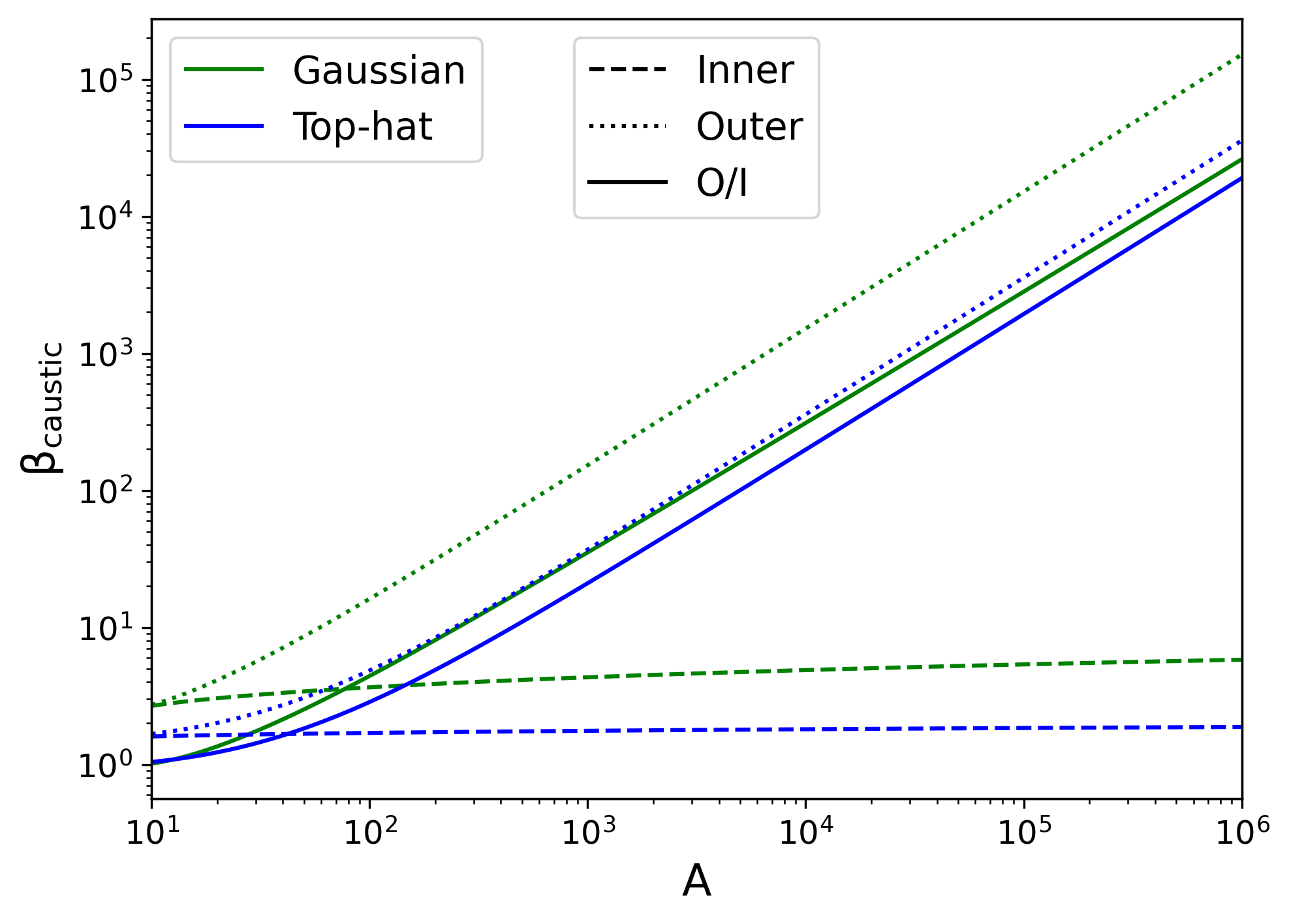}
	\caption{\textbf{The location of caustics $\beta_{\rm caustic}$ as a function of the lens amplitude $A$.} For both the Gaussian and %step-like
    top-hat models, 
    the dimensionless locations $\beta$ of the outer caustics (dotted line) increase linearly with the lens amplitude $A$, whereas those of the inner caustics (dashed line) increase with $A$ only very slightly, that they remain on the order of unity for a large range of $A$. As a consequence, the ratio between the outer and inner caustics (solid line) increases linearly with the lens amplitude $A$, suggesting a much greater range of the three-image zone for a lens with a greater amplitude. 
    The point mass model is not shown because of its lack of outer caustic.
    }
	\label{fig:beta_A}
\end{figure}

Both the Gaussian and the %step-like
top-hat models belong to a class of exponential models \citep{er2018two},
%For exponential model \citep[eq:46]{er2018two}:
\begin{equation}
	\psi_{E} = \frac{A}{h^{2}}\rm{e}^{-\frac{\theta^{h}}{h}} %\sigma^{h}}}.
	\label{eq:ExpModel}
\end{equation}
where $A$ % = A_{0}\theta_{0}^2$ 
is the amplitude of the lens. 
Here and throughout the paper, we have adopted a unit of the typical size of the lens $\sigma$ for the angular locations such as $\theta$ and $\beta$. Correspondingly, the unit for the lens amplitude $A$ is $\sigma^2$.
%we obtain two very different lens shapes by using two values of the index $h$, 
When $h=2$, Eq.\;\ref{eq:ExpModel} gives a Gaussian lens, and when $h=8$, it gives a shape with a flat top and sharp transition at two sides, which we referred to as the %`steplike'
`top-hat' lens (see Figure~\ref{fig:model_shapes}).% is a typical smooth Gaussian distribution structure, and $h=8$ is steplike one with a flat top and sharp transition at two sides, see Figure~\ref{fig:model_shapes}.

% Outer Caustic vs Psi'_max
\begin{figure}[ht!]
	\includegraphics[width=\columnwidth]{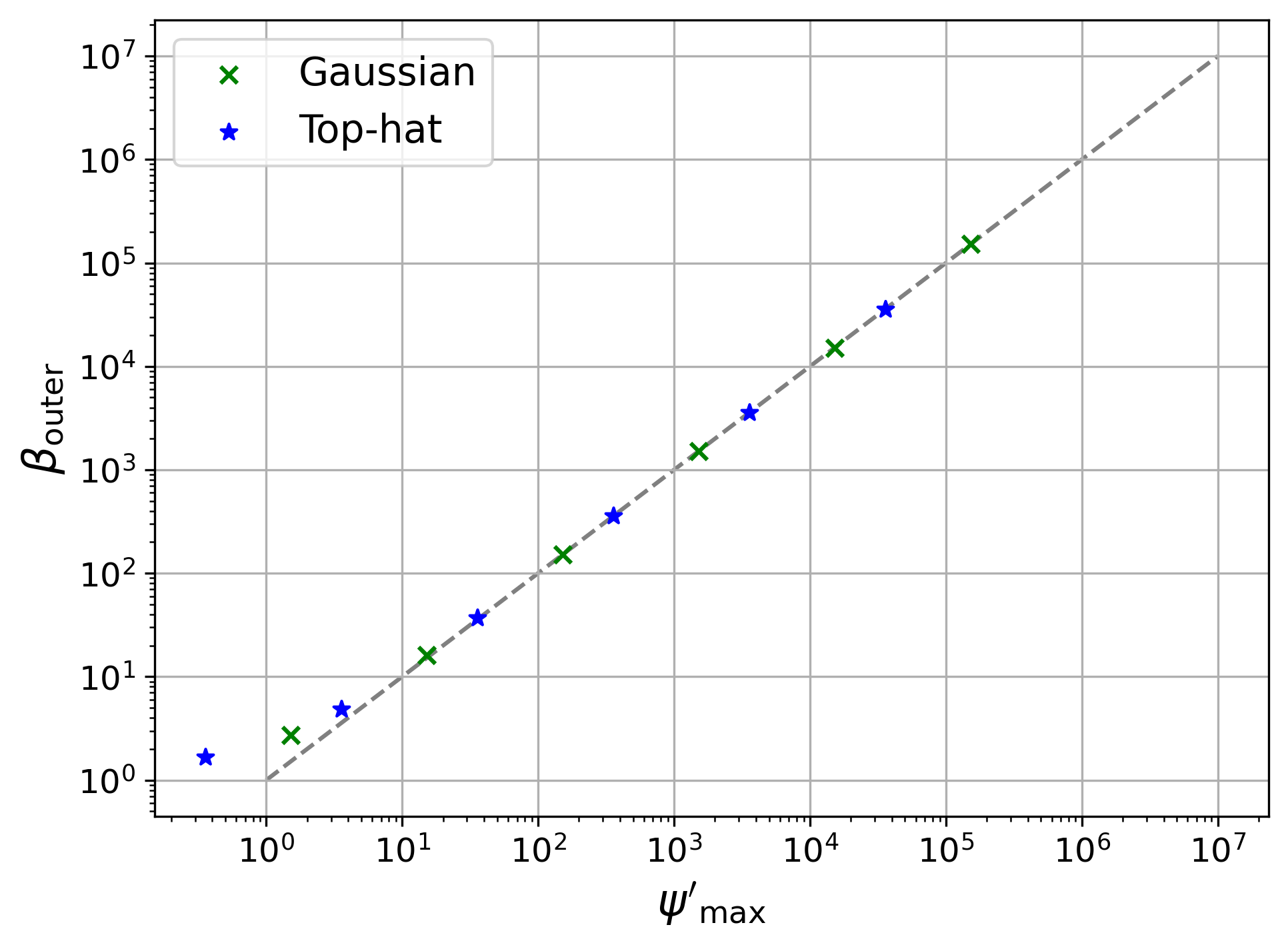}
	\caption{\textbf{The location of the outer caustic $\beta_{\rm outer}$ %as a function 
    versus the maximum gradient of the lens potential $\psi'_{\rm max}$.} The approximate equivalence of the two quantities indicates
      that a measurement of $\beta_{\rm outer}$ can probe the sharpness of the corresponding IISM structure. Shown are Gaussian (green crosses) and %step-like
    top-hat (blue stars) models with amplitudes $A=10, 10^2,...,10^6$ from left to right, respectively. The point mass model also obeys this relation since it has a diverging $\psi'_{\rm max}$ at its centre and thus no finite $\beta_{\rm outer}$. %For both the Gaussian and step-like models, the maximum gradient position stays at almost the same position, so the maximum gradient $\psi'_{\rm max}$ increases linearly with the lens amplitude $A$, thus the dimensionless locations $\beta$ of the outer caustics is directly proportional to (grey dashed line) the maximum gradient for it's linearly relationship with the lens amplitude $A$.
    % whereas those of the inner caustics (dashed line) increase with $A$ only very slightly, that they remain on the order of unity for a large range of $A$. As a consequence, the ratio between the outer and inner caustics (solid line) increases linearly with the lens amplitude $A$, suggesting a much greater range of the three-image zone for a lens with a greater amplitude. The point mass model is not shown because of its lack of outer caustic.
    }
	\label{fig:beta_outer_dpsi_max}
\end{figure}

Point-mass lens %potential 
is a typical model in gravitational lensing. As it reflects a spiky lens potential that is diverging in the centre, we use it here to represent a plasma lens with a very sharp density gradient. Its lensing potential
%, we try the same thing in plasma lensing as a converging lens, and it 
takes the form below:
\begin{equation}
	\psi_{P} = -A{\rm ln|\theta|}
	\label{eq:PointMass}
\end{equation}

For our 1D lenses, %a lensing process, since the plasma lensing process usually exhibits 1-dimensional behaviour,  
the deflection angle $\alpha$ and the magnification $\mu$ can be analytically calculated with %for those potentials~\ref{eq:ExpModel},~\ref{eq:PointMass} 
%can be easily derived by 1D version of deflection angle with equation~\ref{eq:Deflection Angle} and magnification with equation~\ref{eq:Magnification}
\begin{equation}
	\alpha_{X} = \partial_{\theta} \psi_{X}
	\label{eq:DefAngle}
\end{equation}

\begin{equation}
	\mu_{X} = \frac{1}{1-\partial_{\theta}^{2} \psi_{X}}
	\label{eq:Mags}
\end{equation}
where the sub-index $X$ stands for exponential $E$ %, fold $F$ 
or point mass $P$. With condition $\mu_{X} \to \infty$, one can obtain the critical points $\theta_{\rm crit}$ on the image plane. 
Mapping them to the source plane with the lens equation~\ref{eq:lensEq} yields the caustics locations $\beta_{\rm crit}$.
%caustics $\beta$ are just mapping the critical points to the lens plane with lens equation~\ref{eq:lensEq}. 
For a general overdense plasma lens with a positive amplitude $A$, the light is deflected away from its centre, i.e., the lens is ``diverging''. This is the opposite of overdense gravitational lenses, which are converging lenses. For a typical 1D diverging lens, such as the Gaussian lens and the %steplike
top-hat lens, there exist two caustics on each side of the lens centre. We refer to them as the outer and inner caustics, respectively. For the point-mass lens, however, the outer caustic is missing due to the divergence of the gradient of its lensing potential.
%The exponential model usually has two values of caustics on one side of the y-axis, we call them outer and inner caustics, while the point mass lens model only has inner caustics.

% Magnifications
\begin{figure*}[ht!]
    \centering
	\includegraphics[width=2\columnwidth]{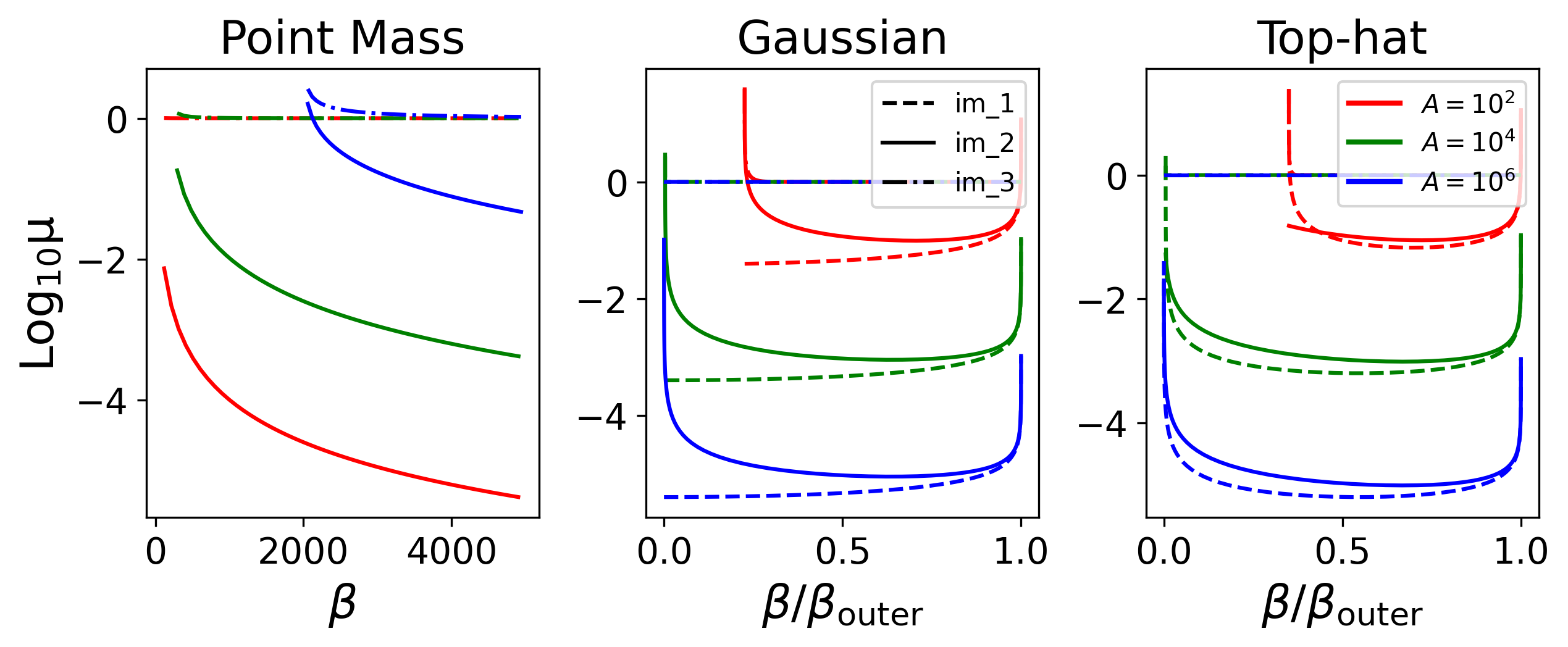}
	\caption{\textbf{Log magnification ${\rm log_{10}\mu}$ as a function of source position $\beta$} for the images in the three-image zone. The magnification of the sub-images created around the lens (solid and dashed lines) depends strongly on the lens amplitude A, indicated by different colours, whereas the magnification of the main image (dash-dotted lines) remains around unity. The point-mass lens has no outer caustic and a different dependence of the sub-image magnification on the lens amplitude due to the singularity in the lens potential. Both the Gaussian and the %step-like
    top-hat models create a pair of sub-images around the lens, which may not be separable due to their close positions. The magnification curves of the sub-image dominant in magnification for both the Gaussian and the %step-like
    top-hat models can be normalised into a unified form (see Fig.\;\ref{fig:mags_exp_fitting}).} %Colour stands for different amplitudes, line style is for different images. For exponential models, their horizontal axis is scaled by the outer caustic and it plots through the entire caustic region from inner to outer caustic, while the point mass models don't have outer caustic, we plot from its inner caustic to 5000, since its outer caustic does not exist, the horizontal axis not have been scaled. For exponential models two extra images could have their signals, but only the brighter one (solid line) could be seen in SS cause the separation between them is too small. The log brightness is obviously concave for all models, which means the brightness increases when the source is close to the caustics. The image brightness changes with amplitude are different among point mass and exponential: Increasing with amplitude for point mass, on the contrary for the exponential model.}
	\label{fig:mags_exp_pm}
\end{figure*}

\section{Results}
\label{sec:Results}

\subsection{Imaging properties and the caustics}

For our given lens models, the lens equation typically yields either one or three solutions, corresponding to %the stationary phase points—i.e., 
the images of the lensing system. 
The lower panel of Figure~\ref{fig:model_shapes} depicts the imaging property of a plasma lens placed away from the pulsar, taking the Gaussian lens model as an example. The black line shows the total phase delay as a function of the angular coordinate $\theta$ with respect to the pulsar angular location $\beta$. Its overall parabolic shape reflects the geometric phase delay, and the small kink at $\theta-\beta=20$ represents the phase delay directly introduced by the plasma lens. The images are, by definition, the extrema of the phase delay curve.

In the depicted case where the plasma lens lies in the ``three-image zone'' between the inner and outer caustics \citep[e.g.,][]{shi2021plasma}, the lensing system has three images (stars in the figure): one near $\theta-\beta=0$ that can be considered as the direct image of the pulsar, and two others around the location of the plasma lens. The two images created at the plasma lens are much fainter than the direct image, as can be seen from the much larger local curvature of the phase delay curve. 
%For a Gaussian lens model, as illustrated by the green curve in the total phase delay plot in the lower panel of Figure~\ref{fig:model_shapes}, the phase delay is generally divided into five regions by the caustics (dashed green line) when the lens moves across the Earth-source line of sight. This profile is symmetric about the y-axis, with two additional extrema—and thus two new images—emerging only when the lens is positioned between the inner and outer caustics. Consequently, the three-image zone lies between these outer and inner caustics. Within this zone, in addition to the main image forming near the source’s location, two fainter images appear around the lens. 
In the context of pulsar scintillation, these additional images manifest as inverted arclets in the secondary spectrum.

Outside the outer caustic, the plasma lens can no longer create local extrema around it. Thus, the additional images vanish. By monitoring the evolution of the inverted arclets along the main parabolic arc, one can in principle observe the creation/vanishing of arclets and constrain the location of the outer caustics corresponding to their underlying plasma lenses. 

Within the inner caustics, the lensing system also has just one image. In this case, however, it is because one of the two images created at the lens has annihilated with the direct image. What is observed is only one faint image created at the plasma lens. This corresponds to the low flux episode created by an ESE. The duration of an ESE thus constrains the size of the inner caustic. In the case of pulsar scintillation, an ESE by a plasma lens that is significantly stronger than the other structures on the phase screen would leave a clear low flux episode on the dynamic spectrum surrounded by the inner caustics \citep[e.g.,][]{shi2021plasma}. In a more commonly observed situation, there exist many plasma lenses underlying the numerous arclets, such as the case shown in Figure~\ref{fig:sec}. When one of these lenses moves within its inner caustic, we expect no distinct feature in the dynamic spectrum. One may observe the disappearance of the arclet near the origin of the secondary spectrum, but the actual observable is complicated by the existence of other lenses of comparable amplitudes and sizes.

%The lower panel of Figure~\ref{fig:model_shapes} illustrates the image formation for a Gaussian lens (green curve) positioned between the inner and outer caustics. Additional images, marked as stars, arise when local extrema form in the phase delay. For a typical three-image scenario, such as with the Gaussian model, three images are produced when the lens lies between the inner and outer caustics. The positions of the caustics (dashed green line in the lower panel of Figure~\ref{fig:model_shapes}) can be determined by setting $\mu_{X} \to \infty$ in Equation~\ref{eq:Mags}. Once the multi-image zones are identified, multiple points within this region can be selected to compute image properties, such as brightness (magnification) and position, for each specified location.

\subsection{Dependence of image formation on the lens amplitude}
Figure~\ref{fig:beta_A} shows the location of the caustics as a function of the lens amplitude $A$. For both the Gaussian and the %step-like
top-hat models, the location of the outer caustic $\beta_{\rm outer}$ increases linearly with $A$ across many orders of magnitude. 
Figure~\ref{fig:beta_outer_dpsi_max} shows an even tighter relation between $\beta_{\rm outer}$ and the maximum gradient of the lens potential, $\psi'_{\rm max}$. In fact, $\beta_{\rm outer}$ almost equals $\psi'_{\rm max}$ in all cases, i.e., for all lens models and various lens amplitudes. This tight relation is not a coincidence: The outer caustic occurs where the second derivative of the lens potential $\psi''$ equals unity (Eq.\;\ref{eq:Mags}), whereas the maximum gradient is located where $\psi''=0$. For a strong lens with $A \gg 1$, it is natural that these two locations are close to each other. To put it into a physical term, the outer caustic approximately marks the maximum deflection angle of a lens.
This sensitivity of the outer caustic location can be used to put a direct constraint on $\psi'_{\rm max}$, which is determined by the maximum gradient of the column density $\Delta N'_{\rm e}$ of the corresponding substructure and can be regarded as a representation of the ``sharpness'' of this plasma lens. 
%Due to the sensitivity of the outer caustic location on the lens amplitude, 
%the former, when observed, can put a direct constraint on the latter, delivering direct information on the density structure on the scintillation screen. 

One caveat of such inferences is highlighted by the point mass lens: a diverging gradient can generate additional extrema anywhere outside the inner caustic, i.e., its outer caustic does not exist, so the three-image zone extends to infinity. 
%\zhu{Thus, an arclet created by a point-mass lens could in principle be observed at an infinite angular separation} between the lens and the pulsar. 
Thus, under the geometrical optics approximation for a point-mass lens, one image formally corresponds to an infinite angular separation (with vanishing flux). While this mathematical limit highlights the model's behavior, the physical observable separation is always finite, fundamentally limited by diffraction and the finite flux threshold for detection.
The underlying reason for the linear dependence of the outer caustic location on the lens amplitude (Figure~\ref{fig:beta_A}) is the dependence of the ability of extrema generation on the gradient of the lens potential, as implied by Figure~\ref{fig:beta_outer_dpsi_max}. For our three lens models with significantly different shapes, it is natural that the model with a sharper gradient has a larger outer caustic location. 

In contrast to the outer caustic, the inner caustic location (dashed lines in Figure~\ref{fig:beta_A}) depends very little on the amplitude of the lens. For any lens shape, the inner caustic is always located at an order of unity in terms of the lens size. Thus, the inner caustic location can be used as an estimate of the lens size. 

Even when the locations of the outer and inner caustics are not captured by observations, multiepoch observations can reveal a range of Doppler values (i.e., pulsar-lens angular separations) where an arclet persists to exist. This ``arclet existence region'' reflects the three-image zone. It puts a lower limit on the ratio of the outer and inner caustics, which then yields a dimensionless constraint on the sharpness of the lens, irrespective of its size. Current observations \citet[e.g.,][]{brisken2010100, sprenger2022doublelens} and Figure\;\ref{fig:sec} typically show that arclets exist over a large range of Doppler values, suggesting a typical value of outer and inner caustics ratio greater, possibly much greater, than 10, indicating a sharp shape of the underlying plasma lenses. 

It is important to note that the lens amplitude for a plasma lens increases quadratically with the observing wavelength, $A\propto \lambda^2$. Correspondingly, the outer caustic for the same density structure in the IISM is located at vastly different locations when observed at different frequencies. Therefore, one may observe the disappearance of an arclet at a certain frequency at which it crosses the outer caustic. Compared to the range of existence of an arclet observed by multiepoch observations that can put a lower limit on the outer caustic location, this frequency dependence of arclet appearance can lead to a precise determination of the outer caustic location, and require only a single-epoch observation. With the recent implementation of ultrawideband receivers in several radio telescopes, it is now possible to look for the outer caustics in a systematic way and use the measurements to constrain the amplitudes of the underlying plasma lenses.

% Table for how observables constrain the model, in which aspects
\begin{deluxetable*}{llll}
%\digitalasset
\tablewidth{0pt}
\tablecaption{Observables and Their Constrains on Lens Model Parameters \label{tab:obs_constraints}}
\tablehead{
\colhead{Observable} & \colhead{Observation Mode} & \colhead{Physical Quantity} & \colhead{Constraint Mechanism}
}
\startdata
Arclet Existence Region & Multiepoch & \makecell[l]{$\bullet$ Caustic Locations ($\beta_{\rm caustic}/\sigma$) 
\\ $\bullet$ Sharpness ($\psi'_{\rm max}/\sigma$) \\ $\bullet$ Amplitude ($A/\sigma^2$)} & \makecell[l]{Arclets exist between caustics;\\ An arclet existing over a larger doppler range \\ suggests a sharper underlying lens. } \\
\hline
\makecell[l]{Frequency-Dependent \\ Arclet Existence/Brightness} & Ultrawideband & \makecell[l]{$\bullet$ Outer Caustic ($\beta_{\rm outer/\sigma}$) \\
$\bullet$ Sharpness ($\psi'_{\rm max}/\sigma$) \\ $\bullet$ Amplitude ($A/\sigma^2$)} & \makecell[l]{Frequency-Ampulitude relation ($A \propto \lambda^{2}$) \\ helps constrain $\beta_{\rm outer}$ and $A$.} \\
\hline
Arclet Brightness Evolution & Multiepoch & \makecell[l]{$\bullet$ Caustic Locations ($\beta_{\rm caustic}/\sigma$) \\ $\bullet$ Sharpness ($\psi'_{\rm max}/\sigma$) \\ $\bullet$ Amplitude ($A/\sigma^2$)} & \makecell[l]{Arclet brightness evolution helps locate caustics.} \\
\hline
ESE \tablenotemark{a} & Single-Epoch & \makecell[l]{$\bullet$ Inner Caustic ($\beta_{\rm inner}/\sigma$) \\ $\bullet$ Size ($\sigma$) } & \makecell[l]{ESE duration constrains inner caustic \\ size that reflects lens size. } \\
\hline
Arclet Pair Separation\tablenotemark{b} & Single-Epoch & $\bullet$ Size ($\sigma$) & \makecell[l]{Separation of arclet pair indicates  lens size.} \\
\enddata
\tablenotetext{a}{Observable only when a distinct, large-amplitude lens transits the pulsar.}
\tablenotetext{b}{Dependent on the identification of arclet pairs.}
%\tablecomments{$^*$ observable only when a distinct, large-amplitude lens transits the pulsar. }
%edition of the {\it Astrophysical Journal}.  A portion is shown here 
%for guidance regarding its form and content. The {\tt\string \digitalasset}\ command highlights the Table title to visually indicate to the reader that there is data associated with this table.}
\end{deluxetable*}

\subsection{Brightness evolution of images generated by substructures}

The brightness of an arclet image evolves as it moves along the main parabolic arc. This brightness evolution remains unmeasured, but it is measurable with multiepoch observations and contains useful information. %, yet its precise distribution remains unmeasured. The single-lens models developed in this paper enable the calculation of this evolving behaviour as arclets traverse the main arc. It was anticipated that the brightness evolution of arclets would mirror the characteristics of the main arc’s brightness distribution, suggesting that unresolved arclets collectively contribute to the main arc. Surprisingly, however, the logarithmic brightness distribution of arclets is found to be concave across all lens shapes (see Figure~\ref{fig:mags_exp_pm}), a qualitative departure from the brightness distribution along the main parabolic arc, as reported by \citet{rickett2021scintillation,brisken2010100}.

Figure\;\ref{fig:mags_exp_pm} shows the evolution of the image brightness as a function of the pulsar angular location $\beta$, with the centre of the lens being placed at $\theta=0$. One can see that the brightnesses of the plasma lensing images (solid and dashed curves) are typically orders of magnitude smaller than those of the direct image (dash-dotted curve around $\log(\nu) = 0$), except in very tiny regions near the inner caustics. This is in accordance with the measured brightness of the extended parabolic arc, which can often extend the whole range between the total flux and the noise limit that spans typically three to four orders of magnitude. 

Out of the pair of images created by the plasma lens, one of them usually dominates in brightness. When the second image is too faint, or when the two images are too close together, one cannot observe an arclet pair but one single arclet. In such a case, one can measure one brightness with respect to the direct image, and can use it to put a constraint on the lens amplitude given its shape. 

An arclet pair that is associated with the same plasma lens has not yet been unambiguously identified. If they are observed, however, their physical separation can serve as a good indicator for the lens size since they are usually produced near the center and the boundary of the lens, respectively (see the lower panel of Figure\;\ref{fig:model_shapes}). Their brightnesses reflect the local curvature of the phase delay curve at the image locations, and thus provide valuable constraints on the shape of the plasma lens. 

\begin{figure}[ht!]
	\includegraphics[width=\columnwidth]{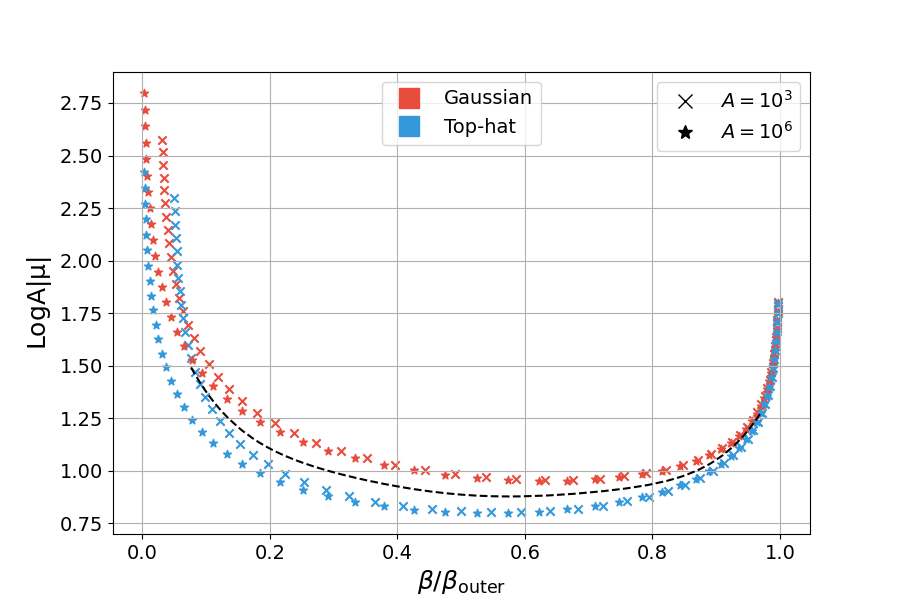}
	\caption{%\textbf{Scaled log magnifications ${\rm log_{10}A\mu}$ as a function of (normalized) source position $\beta/\beta_{\rm outer}$.} Red is for Gaussian and green is for step-like. The marker is for the value of amplitude. The blue line is a polyfit curve to level six. The step-like data shifted a little bit up to about 1.15.
    \textbf{Scaled magnification ${\rm log_{10}(A\mu)}$ as a function of normalized source position $\beta/\beta_{\rm outer}$} for the dominant sub-image in the Gaussian ($h=2$) and %step-like
    top-hat ($h=8$) models. Both models with various lens amplitudes $A$ all exhibit similar functional shape well-described by a polynomial form of ${\rm log_{10}(A\mu)} = 58.25 (\beta / \beta_{\rm outer})^4  - 182.64 (\beta / \beta_{\rm outer})^3 + 231.48 (\beta / \beta_{\rm outer})^2 \\- 151.79 (\beta / \beta_{\rm outer}) + 55.77$  (gray dashed line) away from the caustics.}
    %nearly the same functional form that we fitted with a sixth-order polynomial ${\rm log_{10}A\mu} = 26.08(\beta / \beta_{\rm outer})^4 \\ - 54.90(\beta / \beta_{\rm outer})^3 + 40.92(\beta / \beta_{\rm outer})^2 - 12.87(\beta / \beta_{\rm outer}) + 2.39$ (gray dashed line).}
	\label{fig:mags_exp_fitting}
\end{figure}

The shapes of the brightness evolution exhibit clear similarity for different lens models. When expressed in the log(brightness) - angular location plane, they form a concave curve with peaks near the inner and outer caustics (Figure\;\ref{fig:mags_exp_pm}). This motivates us to normalize the dependencies of the brightness evolution on the lens model and the lens amplitude, and the result is shown in Figure\;\ref{fig:mags_exp_fitting}. Expressed in terms of $\log(A\nu) - \beta/\beta_{\rm outer}$, the brightness evolution for both the Gaussian and the %step-like
top-hat models and various lens amplitudes $A=10^2, 10^4,$ and $10^6$ lie nearly on the same curve. A polynomial fitting for this curve yields  ${\rm log_{10}A\mu} = 26.08(\beta / \beta_{\rm outer})^4 \\ - 54.90(\beta / \beta_{\rm outer})^3 + 40.92(\beta / \beta_{\rm outer})^2 - 12.87(\beta / \beta_{\rm outer}) + 2.39$.
This suggests that the shape of the brightness evolution is insensitive to the lens model or lens amplitude, but a measurement of a small section of the brightness evolution may be used to infer the relative location of the corresponding pulsar location with respect to the outer caustic. This can serve as another effective way of inferring the caustic location.

% Magnifications fitting

\subsection{Observation prospects of substructure properties}

Table~\ref{tab:obs_constraints} summarizes the key observable features from the secondary spectrum of pulsar scintillation and their implications for the plasma lens models. It details five observables -- arclet existence region, frequency-dependent arclet existence/brightness, arclet brightness evolution, ESE, and arclet pair separation. When observed via multiepoch, ultrawideband, or single-epoch mode, they can help infer corresponding physical quantities e.g.,, caustic locations $ \beta_{\rm caustic}/\sigma$, sharpness $\psi'_{\rm max}$, dimensionless lens amplitude $A/\sigma^2$, and lens size $ \sigma $. Note that we have put the unit $\sigma$ back in this table for clarity. %These are measured as follows: multi-epoch observation tracks the caustic location and records the brightness evolution; ultra-Wide Band detects frequency-amplitude relations ($ A \propto \lambda^2$ ) to refine outer caustic and $ A $; Single-Epoch ESE duration sets inner caustic size; and Single-Epoch pair separation indicates lens size. However, observing ESEs and arclet pair separations is difficult, as they rely on the rare occurrence of large-amplitude lens transits and the precise identification of arclet pairs, respectively.
Among them, using a single-epoch observation of a pulsar ESE event to constrain the inner caustic and the lens size is most straightforward to perform. However, it relies on the chance detection of a rare transit of a large-amplitude lens. Using arclet separation in single-epoch observations to constrain lens sizes is also challenging, as it relies on the clean identification of arclet pairs. 
Determining the arclet existence region and arclet brightness evolution using multiepoch observations can constrain the caustic locations and the lens amplitude. This has not been systematically pursued, likely due to its stringent requirements on telescope time. With ultrawideband receivers becoming available \citep{hobbs2020ultrawide}, 
%(cite https://arxiv.org/abs/1911.00656), 
observing frequency-dependent arclet existence and brightness is well-suited for a systematic effort to constrain lens properties such as the outer caustic location and lens amplitude.

\begin{figure}[ht!]
	\includegraphics[width=\columnwidth]{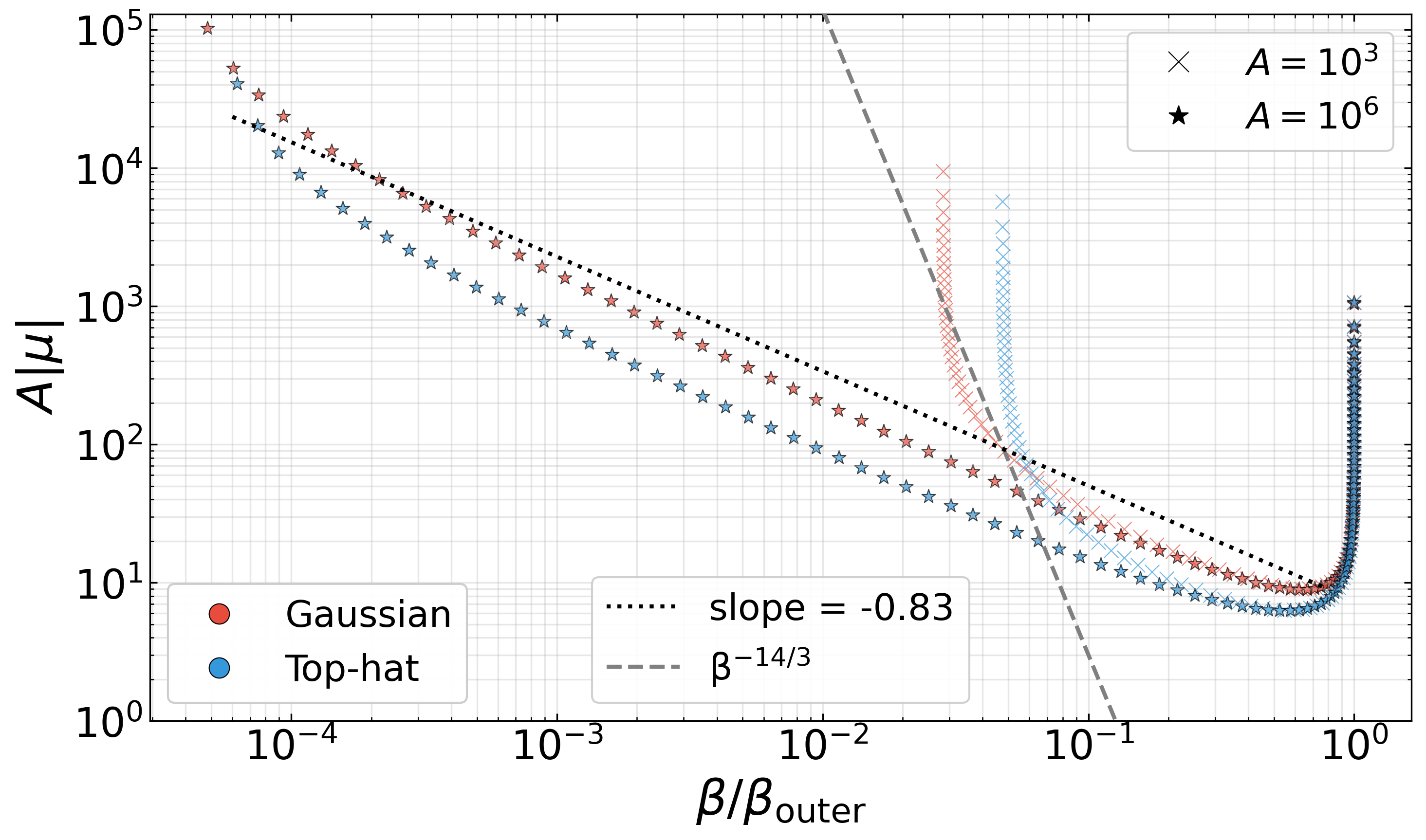}
	\caption{%\textbf{Scaled log magnifications ${\rm log_{10}A\mu}$ as a function of (normalized) source position $\beta/\beta_{\rm outer}$.} Red is for Gaussian and green is for step-like. The marker is for the value of amplitude. The blue line is a polyfit curve to level six. The step-like data shifted a little bit up to about 1.15.
    %\textbf{Scaled magnification ${\rm A\mu}$ as a function of normalized source position $\beta/\beta_{\rm outer}$ in log-log space.} Comparing the slope of fitting (black dotted line) the dominant sub-image in the Gaussian ($h=2$) and top-hat ($h=8$) models with $\beta^{-14/3}$ (dashed grey line).}
    \textbf{Magnification of the dominant sub-image scales with the source position to a power of less than -1 in regions well inside the caustics.} This scaling slope is much shallower than the $\beta^{-14/3}$ slope (dashed grey line) characteristic of the overall brightness distribution of all scattered images observed as the main parabolic arc.}    
	\label{fig:mags_exp_fitting_loglog}
\end{figure}

Most of the observables in the table rely on the detection of the arclet in the first place. Here, by an ``arclet'', we mean an observable on the secondary power spectrum that is associated with the lensed image created by a discrete structure. It usually takes the form of an inverted arclet in the case of strong scintillation, but is also measurable as a discrete flux concentration on the main parabolic arc in the case of weak scintillation. 
The key criterion of the detection of an arclet is that its flux exceeds that of the main parablic arc created by an ensemble of scattered images. In Fig.\;\ref{fig:mags_exp_fitting_loglog} we compare the brightness of an arclet as it moves across the main parabolic arc with the overall brightness distribution of the main parabolic arc. The brightness of the main parabolic arc decreases steeply with the angular separation from the pulsar, typically with a power-law slope of about $-14/3$ characteristic of Kolmogorov turbulence \citep[see e.g.,][]{ocker2024pulsar}. In comparison, the brightness of an arclet, when well inside of the three-image zone bounded by the inner- and outer-caustics, scales with the angular separation to a relatively shallow power of less than $-1$. 
This means that, whereas an extremely bright arclet can be observed at all locations within its existence region, an arclet not bright enough can distinct itself from the main parabolic arc only at large angular separations. For such an arclet, observables related to its outer caustic, like its maximum existence separation, frequency-dependent existence, and brightness evolution at large angular separations, are key to probe its properties.

\section{Conclusion}
\label{sec:Conclusions}
This investigation has illustrated the role of pulsar scintillation in probing the microphysical structures of the IISM through modeling. We construct three distinct lens models (point mass, Gaussian, and %step-like
top-hat) that would represent different astrophysical origins, and analyze the lensing properties of these substructure lenses. The model results can be directly compared to observations since the substructure images can be measured as inverted arclets on the pulsar secondary spectra.

We have demonstrated the outer caustic's sensitivity to the lens amplitude and observing frequency, which is well-suited for constraining the maximum column density gradient, i.e.,  the ``sharpness'' of IISM density structures. %with unprecedented precision.
The evolution of the logarithmic brightnesses of substructure images forms a concave function of the pulsar-lens angular separation. This is in contrast with the convex main parabolic arc brightness distribution, highlighting the complementarity of individual arclet studies to the latter, statistical studies. Employing the quasi-universal shape of the scaled brightness-location function, one can estimate the caustic locations based on relatively few arclet brightness measurements.

The detailed dependencies of the observables on the lens properties also underlie the utility of pulsar scintillation as a suitable probe for the IISM substructure.
The constraints derived from arclet existence region, frequency-dependent arclet existence and brightness, and outer-to-inner caustic ratios offer insights into the distinct arclet observables induced by different lens types, providing ways to probe the lens properties and thus the unknown plasma density structures in the IISM.

Future work could focus on utilizing ultrawideband receivers to determine the frequency points of outer caustic disappearance for precise localization, employing multiepoch observations to track caustic positions and inverted arclet brightness evolution, and exploring databases of high-precision single-epoch observations for pulsar ESE events and cases with cleanly identifiable arclet pairs.
Such advancements will further illuminate the astrophysical processes governing IISM dynamics, paving the way for refined models of interstellar turbulence and plasma distributions.

\begin{acknowledgments}
This work is supported by NSFC grant No. 12373025. %We thank the organizers and speakers of the “Gravity meets Plasma” workshop held in Yunnan University in summer 2019 for introducing this field to us.
This work made use of the data from FAST (Five-hundred-meter Aperture Spherical radio Telescope).  FAST is a Chinese national mega-science facility, operated by National Astronomical Observatories, Chinese Academy of Sciences.
\end{acknowledgments}


%% For this sample we use BibTeX plus aasjournalv7.bst to generate the

\bibliography{P_lensing}{}
\bibliographystyle{aasjournalv7}



\end{document}